\documentstyle[twocolumn,prb,aps,epsfig]{revtex}

\begin{document}

\title{Spin glass like transition in a highly concentrated Fe-C nanoparticle system}

\author{P. J{\"o}nsson, P. Svedlindh and P. Nordblad}
\address{Department of Materials Science, Uppsala University\\
        Box 534, SE-751 21 Uppsala, Sweden}

\author{M. F. Hansen}
\address{Department of Physics, Building 307, Technical University of Denmark\\
DK-2800 Lyngby, Denmark}

\date{\today}

\maketitle

\begin{abstract}
A highly concentrated (17 vol.\%) Fe-C nano-particle system, with a narrow size distribution $d = 5.4\pm 0.4$~nm, has been investigated using magnetic ac susceptibility measurements covering a wide range of frequencies (17 mHz - 170 Hz). A dynamic scaling analysis gives evidence for a phase transition to a low temperature spin-glass-like phase. The critical exponents associated with the transition are $z\nu = 10.5 \pm 2$ and $\beta = 1.1 \pm 0.2$. 
The reason why the scaling analysis works for this sample, while it may not work for other samples exhibiting collective behavior as evidenced by aging phenomena, is that the single particle contribution to $\chi''$ is vanishingly small for $T>T_g$ and hence all slow dynamics is due to collective behavior. 
This criterion can only be fulfilled for a highly concentrated nano-particle sample with a narrow size distribution. 

\end{abstract}

\section{Article}

Dense nano-particle systems have been shown to exhibit collective behavior, as evidenced by aging, and non-equilibrium effects similar to spin glasses.\cite{Dormann,Mamiya,PJ} Whether that collective behavior is associated with a ``true'' spin glass phase transition is still controversial. 
In this paper, we show a dynamic scaling analysis to a spin glass transition for a highly concentrated Fe-C nano-particle sample with a narrow particle size distribution. We also discuss why the scaling analysis may not indicate a phase transition for samples with less interactions or wider size distributions.

The sample consisted of ferromagnetic nanoparticles of amorphous
$\mathrm{Fe_{1-x}C_x}$ (x$\approx$0.22), with an average particle size of $5.4\pm0.4$ nm, prepared by the method described by Van Wonterghem et al. in Ref. \onlinecite{Wonterghem}.
The sample was studied in the frozen state and contained 17 vol\% of particles. The ac-susceptibility measurements were performed in a non-commercial SQUID magnetometer for frequencies in the range $\omega/2\pi =$ 17~mHz - 170~Hz. 
Fig. 1 shows $\chi'(T)$ and $\chi''(T)$ at different frequencies.

A sample that exhibit a spin glass transition will show critical slowing down, and hence the characteristic relaxation time $\tau$ diverges at the transition temperature according to
\begin{equation}
\tau = \tau_* (T/T_g-1)^{-z\nu}, \qquad T>T_g
\end{equation}
where $T_g$ is the transition temperature, $\tau_*$ is related to the relaxation time of the individual particle magnetic moments, and $z\nu$ is a critical exponent.
We extracted the freezing temperature $T_f$, associated with a relaxation time ($\tau = 1/\omega$), from the out-of-phase component of the ac-susceptibility as $\chi''(T_f) = \frac{1}{7} \chi''_{max}$ with $T_f >T_g$.
We also tried other criteria to make sure that the choice of criterion is not significantly influencing the results of the critical slowing down analysis.
The critical slowing down analysis gives $z\nu = 10.5 \pm 2$, $\tau_* = 2.2\times10^{-8}$~s, and $T_g = 49.5$~K (see Fig. 2).

We also performed a full scaling of $\chi''$ according to,
\begin{equation}
\chi''/\chi_{eq} = (T/T_g-1)^\beta H(\omega \tau), \qquad T>T_g
\end{equation}
and found data collapse to a single function $H(\omega \tau)$, for different frequencies, using $\beta = 1.1 \pm 0.2$ (see Fig. 3).
The value of $z\nu$ compares quite well with values found for spin glasses with long range interactions (RKKY), while the value of $\beta$ is slightly larger than typical spin glass values, but is consistent with the value of $\beta = 1.2 \pm 0.1$ found by Jonsson et al. \cite{Jonsson} for an interacting nanoparticle sample by a static scaling analysis.

The dynamic scaling analysis will only reveal a phase transition if the single particle contribution to $\chi''$ is vanishingly small for $T>T_g$, i.e. all slow dynamics is due to collective behavior. Two criteria have to be fulfilled for this to be possible; i) the interparticle interactions need to be strong and ii) the particle size distribution needs to be narrow.  
If we compare the sample used for the scaling analysis with the same sample but much more dilute (0.05 vol.\%) we can see that the out-of-phase component is almost zero for $T>T_g$ (see inset of Fig. 1). 
We conclude that the concentrated sample is appropriate to use for scaling analysis.

Financial support from NFR is acknowledged.

\begin{figure}[htb]
\centerline{\epsfig{figure=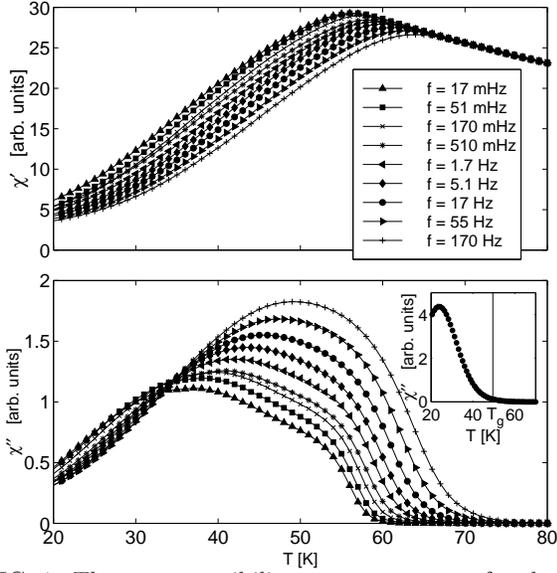,width=7.5cm}}
\caption[]{The ac-susceptibility vs. temperature for the concentrated sample (17 vol.\%). Inset: The ac-susceptibility for the dilute sample (0.05 vol.\%), $f=125$~Hz.}
\label{susc}
\end{figure}

\begin{figure}[htb]
\centerline{\epsfig{figure=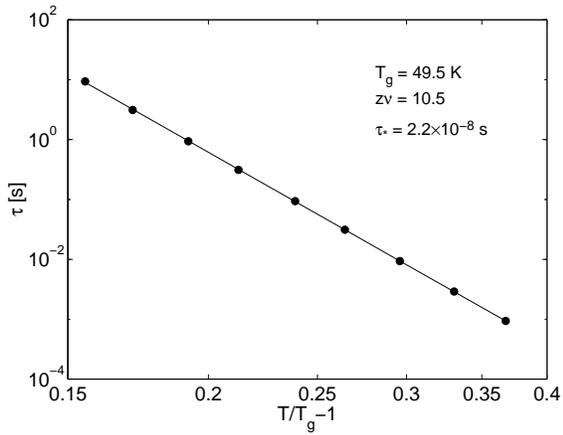,width=7.5cm}}
\caption[]{Critical slowing down analysis.}
\end{figure}

\begin{figure}[htb]
\centerline{\epsfig{figure=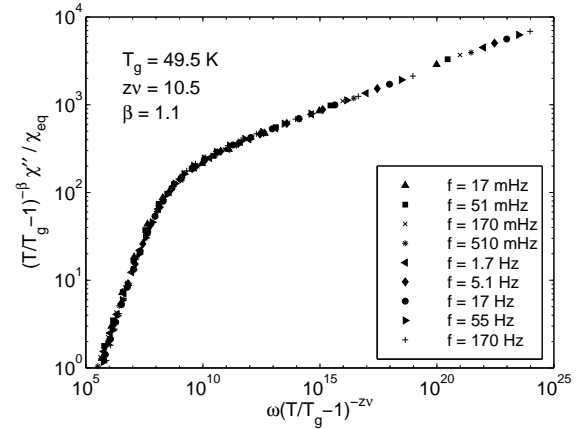,width=7.5cm}}
\caption[]{ Scaling of $\chi''(T,\omega)/\chi_0 = (T/T_g-1)^\beta$ using data for $T>T_g$.}
\end{figure}

\end{document}